\documentclass[fleqn,usenatbib]{aa}

\DeclareRobustCommand{\VAN}[3]{#2}
\let\VANthebibliography\thebibliography
\def\thebibliography{\DeclareRobustCommand{\VAN}[3]{##3}\VANthebibliography}

\usepackage{comment}
\usepackage{graphicx,amsmath,amsfonts,amssymb,slashed,float}

\usepackage{hyperref}
\hypersetup{breaklinks=true,colorlinks=true,linkcolor=blue,urlcolor=blue,citecolor=blue}

\usepackage[normalem]{ulem}
\usepackage{bbold,wasysym}
\usepackage{array,multirow}
\usepackage[utf8]{inputenc}
\usepackage[usenames,dvipsnames]{xcolor} 
\usepackage{soul}
\usepackage[pscoord]{eso-pic}
\usepackage{makecell}
\usepackage{placeins}
\usepackage{subfigure}
\usepackage{orcidlink}
\usepackage{listings}
\usepackage{xcolor}
\usepackage{comment}

\usepackage{newtxtext,newtxmath}
\usepackage[T1]{fontenc}

\newcommand{\LCDM}{$\Lambda$CDM}

\newcommand*{\TT}{\ensuremath{T\!T}}
\newcommand*{\EE}{\ensuremath{E\!E}}
\newcommand*{\TE}{\ensuremath{T\!E}}

\newcommand*{\pp}{\ensuremath{\phi\phi}}

\newcommand*{\TTTEEE}{\ensuremath{T\!T/T\!E/E\!E}}

\newcommand*{\planck}{\textit{Planck}}

\newcommand*{\candl}{\texttt{candl}}

\newcommand{\RNum}[1]{\MakeUppercase{\romannumeral #1}}

\definecolor{codegreen}{rgb}{0,0.6,0}
\definecolor{codegray}{rgb}{0.5,0.5,0.5}
\definecolor{codepurple}{rgb}{0.58,0,0.82}
\definecolor{backcolour}{rgb}{0.95,0.95,0.92}

\lstdefinestyle{mystyle}{
    backgroundcolor=\color{backcolour},   
    commentstyle=\color{codegreen},
    keywordstyle=\color{magenta},
    numberstyle=\tiny\color{codegray},
    stringstyle=\color{codepurple},
    basicstyle=\ttfamily\footnotesize,
    breakatwhitespace=false,    
    columns=fullflexible,
    breaklines=true,                 
    captionpos=b,                    
    keepspaces=true,                 
    numbersep=5pt,                  
    showspaces=false,                
    showstringspaces=false,
    showtabs=false,                  
    tabsize=2
}

\newcommand{\skipt}[1]{}

\definecolor{RoyalBlue}{rgb}{0.25,.41,.88}
\definecolor{celestialblue}{rgb}{0.29, 0.59, 0.82}

\setstcolor{Blue}

\begin{document}

\title{$\texttt{candl}$: cosmic microwave background analysis with a differentiable likelihood\thanks{\candl{} is available at \url{https://github.com/Lbalkenhol/candl}}}
\authorrunning{L. Balkenhol et al.}


\author{L. Balkenhol\,\orcidlink{0000-0001-6899-1873},
      \inst{1}
      C. Trendafilova,
      \inst{2}
      K. Benabed,
      \inst{1}
      \and
      S. Galli
      \inst{1}
      }

\institute{
Sorbonne Universit\'{e}, CNRS, UMR 7095, Institut d'Astrophysique de Paris, 98 bis bd Arago, 75014 Paris, France\\
E-mail: \href{mailto:lennart.balkenhol@iap.fr}{lennart.balkenhol@iap.fr}
\and
Center for AstroPhysical Surveys, National Center for Supercomputing Applications, University of Illinois Urbana-Champaign, Urbana, IL, 61801, USA
}

\date{Received XXX; accepted YYYY}

\abstract{
We present \candl{}, an automatically differentiable python likelihood for analysing cosmic microwave background (CMB) power spectrum measurements.
\candl{} is powered by JAX, which makes it fast and easy to calculate derivatives of the likelihood.
This facilitates, for example, robust Fisher matrices without finite-difference methods.
We show the benefits of \candl{} through a series of example calculations, covering forecasting, robustness tests, and gradient-based Markov chain Monte Carlo sampling.
These also include optimising the band power bin width to minimise parameter errors of a realistic mock data set.
Moreover, we calculate the correlation of parameter constraints from correlated and partially overlapping subsets of the SPT-3G 2018 \TTTEEE{} data release.
In a traditional analysis framework, these tasks are slow and require careful fine-tuning to obtain stable results.
As such, a fully differentiable pipeline allows for a higher level of scrutiny;
we argue that this is the paradigm shift required to leverage incoming data from ground-based experiments, which will significantly improve the cosmological parameter constraints from the \planck{} mission.
\candl{} comes with the latest primary and lensing power spectrum data from the South Pole Telescope and Atacama Cosmology Telescope collaborations and will be used as part of the upcoming SPT-3G \TTTEEE{} and \pp{} data releases.
Along with the core code, we release a series of auxiliary tools, which simplify common analysis tasks and interface the likelihood with other cosmological software.
\candl{} is pip-installable and publicly available on Github.
}


\maketitle



\section{Introduction}\label{sec:intro}

Cosmic microwave background (CMB) observations are a corner stone of modern cosmology.
Measurements of the temperature and polarisation power spectrum of the CMB allow us to determine the free parameters of the $\Lambda$ cold-dark matter (\LCDM{}) model with high precision and tightly limit the space of possible extensions \citep{planck18-6, aiola20, balkenhol23}.
While the \planck{} satellite has mapped the foreground-subdominant scales of the temperature power spectrum to the sample variance limit \citep{planck18-5}, contemporary ground-based experiments, such as the South Pole Telescope (SPT) \citep{carlstrom11}, the Atacama Cosmology Telescope (ACT) \citep{kosowsky03}, and the Simons Observatory (SO) \citep{ade19} complement the satellite data and map the CMB's polarisation anisotropies on intermediate and small angular scales with unseen precision.
Data collected by each of these telescopes is expected to come close to or exceed the parameter constraints of the \planck{} data set \citep{ade19} \citep[based on the lensing data presented in][]{madhavacheril23, qu23} (Prabhu et al. in preparation).

This high-precision data demands a high level of scrutiny.
Any potential claim of a detection of physics beyond the Standard Model can only be embraced with confidence by the scientific community, if the analysis is demonstrably robust.
The growing complexity of CMB analyses obliges a plethora of tests of alternate analysis choices to assert that conclusions remain unchanged.
Moreover, complex forecasting tasks need to be carried out for the survey design and planning process of forthcoming CMB experiments, such as CMB-S4 \citep{abazajian16}.
Ergo, fast, flexible, and robust tools for the analysis of CMB data are indispensable.

In this work, we present \candl{}: an automatically differentiable, stand-alone python likelihood designed for the analysis of high-precision primary CMB (\TTTEEE{}) and lensing power spectrum (\pp{}) measurements.
Calculating derivatives of the likelihood is easy and robust with \candl{};
this ready access to Fisher matrices trivialised a plethora of common analysis tasks, including forecasting, propagating biases from band powers to parameters, and finding the maximum of the likelihood surface, which corresponds to the best-fit point.
We show explicit examples of these tasks in this work, including optimising the band power bin width of a realistic mock data set to minimise parameter errors and calculating the correlation of parameter constraints from correlated and partially overlapping subsets of the SPT-3G 2018 \TTTEEE{} data set \citep{balkenhol23}.
\candl{} interfaces seamlessly with established cosmological software, such as Cobaya \citep{torrado21}, MontePython \citep{brinkmann19, audren13}, CAMB \citep{lewis00}, CLASS \citep{blas11}, and CosmoPower \citep{spuriomancini22}.
The likelihood code will form a part of the analyses pipelines of the upcoming SPT-3G \TTTEEE{} and \pp{} data releases.

\candl{} likelihoods are differentiable thanks to JAX \citep{jax18}.\footnote{\url{https://github.com/google/jax}}
JAX is a python-library for high-performance array computations and is as such well-suited for CMB likelihoods.
JAX offers just-in-time compilation, GPU optimisation, and automatic vectorisation.
Crucially, JAX features an automatic differentiation algorithm, which can be used to calculate derivatives of complicated functions quickly and robustly.
Due to the aforementioned advantages, JAX has seen increasing use in cosmology \citep{campagne23, piras23, kvasiuk23}.
We refer the reader to \citet{campagne23} for an excellent introduction to differentiable likelihoods and the wider potential JAX holds for the field of Cosmology.
\citet{campagne23} touch on many of the techniques discussed in this work and pedagogically show the benefits of a differentiable framework.
We understand \candl{} as a related effort to JAX-COSMO \citep{campagne23}, designed specifically for the case of CMB data analysis and its particular requirements.
Still, \candl{} can be used without JAX for traditional inference.

This work is structured as follows.
In \S\ref{sec:background} we discuss relevant background information.
We provide an overview of the core likelihood code and additional aspects of the released material in \S\ref{sec:overview}.
In \S\ref{sec:diff_like} we highlight applications of the differentiability aspect of \candl{}.
We show how to get started using the code in \S\ref{sec:trad} before concluding in \S\ref{sec:conclusions}.

\section{Background}\label{sec:background}

The likelihood $\mathcal{L}$ is the probability of measuring the data band powers at hand given a certain model.
In the simplest case, the likelihood is Gaussian:
\begin{equation}
\begin{aligned}
    -2\log{\mathcal{L}(\theta | \hat{D})} = &\left\{\hat{D} - \mathcal{T}\left[D^{\mathrm{CMB}}(\theta), \theta \right]\right\}^T C^{-1}\\
    &\left\{\hat{D} - \mathcal{T}\left[D^{\mathrm{CMB}}(\theta), \theta \right]\right\},
\end{aligned}
\label{eq:like}
\end{equation}
where $\hat{D}$ are the measured band powers with covariance $C$, $\theta$ are the model parameters (cosmological and nuisance), $D^{\mathrm{CMB}}$ is the CMB power spectrum, and $\mathcal{T}$ represents the 'data model', that is the transformation of the CMB power spectra such that they can be compared to the data, for example foreground contamination, binning, calibration.
Since the covariance matrix is constant we ignore the contribution of its determinant \citep{percival06, hamimeche08, martina20}.

In practice, the calculation of Equation \ref{eq:like} is typically split into two parts.
First, the 'theory code' calculates $D^{\mathrm{CMB}}(\theta)$, that is the CMB spectra for a given set of cosmological parameters; the theory code may be a Boltzmann solver such as CAMB \citep{lewis00} or CLASS \citep{blas11}, or a surrogate emulator such as CosmoPower \citep{spuriomancini22}, Capse.jl \citep{bonici23}, or the algorithm put forward by \citet{guenther23}.
Second, the 'likelihood code', or simply 'likelihood' for short, transforms the model spectra and completes the calculation.
\candl{} implements this part.

\section{Overview}\label{sec:overview}

\candl{} is a python-based, JAX-powered likelihood: its likelihoods are automatically differentiable, offering ready access to robust derivatives and Fisher matrices.
The code is self-contained, such that one can easily interact with a given data set and interface \candl{} with other cosmology software smoothly.
\candl{} can be installed via \lstinline{pip install candl-like};
detailed instructions are provided on the GitHub repository\footnote{\url{https://github.com/Lbalkenhol/candl}}.

We have made the following public CMB data sets available with the release of \candl{}:
\begin{enumerate}
    \item SPT-3G 2018 \TTTEEE{} \citep{balkenhol23},
    \item SPT-3G 2018 \pp{} \citep{pan23},
    \item ACT DR4 \TTTEEE{} \citep{aiola20, choi20},\footnote{The frequency-compressed, foreground marginalised, 'CMB-only' likelihood.}
    \item ACT DR6 \pp{} \citep{madhavacheril23, qu23}.
\end{enumerate}
The reimplementations of the above public data sets we provide match the log likelihood value of their corresponding official releases to within $\lesssim 10^{-4}\%$ given the same input CMB power spectra and nuisance parameter values.
The upcoming SPT \TTTEEE{} and \pp{} releases will use \candl{} and we are committing to including other data sets as they become available to provide a unified framework to access and use popular CMB data.

Along with the core likelihood code, we supply a suite of auxiliary tools designed to help the user carry out common analysis tasks.
Furthermore, we release a series of example notebooks that showcase different applications of the likelihood.
These resources can be found on the GitHub repository and include:
\begin{itemize}
    \item A showcase of the automatic-differentiation features of the likelihood, including the calculation of Fisher matrices and use of gradient-based minimisation and sampling algorithms.
    \item An interface between the likelihood code and the commonly-used Markov chain Monte Carlo (MCMC) packages Cobaya \citep{torrado21} and MontePython \citep{brinkmann19, audren13}, as well as an interface with the theory codes CAMB \citep{lewis00}, CLASS \citep{blas11}, CosmoPower \citep{spuriomancini22}, CosmoPower-JAX \citep{piras23}, and Capse.jl \citep{bonici23} independent of an MCMC sampler.
    \item Tools for the generation of mock data, the construction of a minimum-variance combination of the data from a multi-frequency version, and various inter-frequency consistency tests.
\end{itemize}

\section{Applications of a differentiable inference pipeline}\label{sec:diff_like}

\begin{figure*}
    \centering
    \includegraphics[width=2.0\columnwidth]{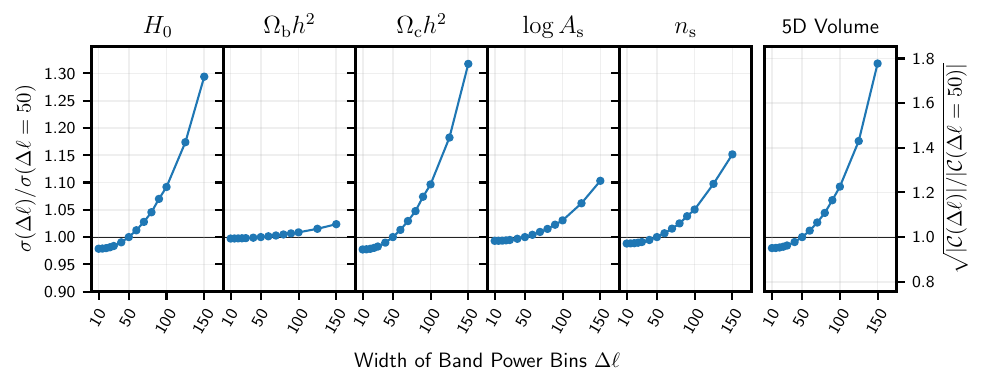}
    \caption{Relative size of error bars of \LCDM{} parameters ($\Omega_{\mathrm{c}} h^2, \Omega_{\mathrm{b}} h^2, H_0, n_{\mathrm{s}}, \log{A_{\mathrm{s}}}$) for different bin sizes $\Delta\ell$ with respect to $\Delta\ell=50$ for a mock CMB data set.
    In the rightmost panel, we show the determinant of the parameter covariance $\mathcal{C}$, a proxy for the five-dimensional volume, using a different scale for the vertical axis.
    We note that the parameter that is affected the most by the broadening of bins is $H_0$, which relies on an accurate measurement of the acoustic peak positions.
    Due to the quadratic relationship between $\Delta\ell$ and $\sigma$, bin sizes smaller than $\Delta\ell=50$ yield a negligible improvement on marginalised parameter errors ($\leq 3\%$).
    Forecasts are fast and easy with \candl{}, enabling thorough analysis optimisation.}
    \label{fig:binning}
\end{figure*}

In this section, we highlight a range of applications enabled by the pairing of \candl{} with a differentiable theory code.
This creates a fully differentiable pipeline and provides easy access to derivatives of the likelihood with respect to cosmological and nuisance parameters, trivialising for example the calculation of robust Fisher matrices without the need for finite-difference methods.

As \citet{campagne23} discuss, the most common solution for a differentiable theory code is to use a differentiable emulator \citep[e.g.][]{spuriomancini22, piras23, bonici23, nygaard23} of spectra computed with the Boltzmann solvers CAMB \citep{lewis99} or CLASS \citep{blas11}.
We note that fully differentiable Boltzmann solvers have recently been developed in JAX \citep{hahn23} and Julia \citep{ruizzapatero23} (\href{https://github.com/xzackli/Bolt.jl}{BOLT.jl}, Z. Li and J. Sullivan in preparation).
We anticipate a straightforward interface between \candl{} and the JAX-based work of \citet{hahn23}, similar to the one we have realised with CAMB and CLASS.
Fundamentally, the log likelihood evaluation in \candl{} is a function call, which requires a dictionary of parameter values as its argument;
this function can be concatenated with the evaluation call of any Boltzmann solver with a minimal interface performing any necessary data format conversions.
When coupling \candl{} with JAX-based theory codes, the full advantages of the library remain available, whereas an inter-language interface (e.g. with Julia codes) may surrender these.

In the examples below, we use the CosmoPower CMB primary power spectrum emulators trained on high-accuracy CAMB spectra developed and used in \citet{balkenhol23} together with CosmoPower-JAX \citep{piras23}.\footnote{These trained models are publicly available: \url{https://github.com/alessiospuriomancini/cosmopower}}
In using emulators, one must be careful to match the accuracy of the emulator to the requirements set by the data set; we have encountered no such issues with the chosen models and data sets for the purpose of these examples.

\subsection{Forecasting}\label{sec:forecast}

CMB power spectrum measurements are typically binned into band powers for two reasons:
first, in order to limit the computational cost of the likelihood evaluation by keeping the data vector small and second, because masking leads to mode-coupling, which prohibits the reconstruction of all individual modes.
However, excessively coarse bins risk losing sensitivity to fine-grained features of the power spectrum, thus unnecessarily widening constraints on cosmological parameters.
Therefore, we wish to optimise the bin width $\Delta\ell$ and find the point of diminishing returns.\footnote{Practically, the largest array size feasible for computations is to be decided with the available hardware and the number of spectra in mind.}

In this example, we consider a $150\,\mathrm{GHz}$ \TTTEEE{} mock data set covering $\ell=300-4000$.
The covariance matrix contains realistic noise contributions and off-diagonal correlations based on the SPT-3G 2018 \TTTEEE{} data set; we use a rectangular survey mask covering approximately $4\%$ of the sky and similar noise curves though with a white noise level of $5\mu K\,\mathrm{arcmin}$.
We use the foreground model from the SPT-3G 2018 \TTTEEE{} data set\footnote{We do not include the CIB-tSZ term due to its non-differentiable form.} and the same \planck{}-based prior on the optical depth to reionisation $\tau$ \citep{planck18-6}.
For this data set, we consider 14 binning options in the range $\Delta\ell=10-150$, allowing bins to exceed the maximum multipole moment if necessary.
This has an insignificant effect on our results due to the negligible cosmological information at these scales and the noise in the data.

To accomplish this forecasting task, we instantiate a \candl{} likelihood for each binning option and calculate the Fisher matrix \citep{heavens16} using CosmoPower-JAX and the supplied tools.
We plot expected parameter errors on the cosmological parameters $\Omega_{\mathrm{c}} h^2$, $\Omega_{\mathrm{b}} h^2$, $H_0$, $n_{\mathrm{s}}$, and $\log{A_{\mathrm{s}}}$ in Figure \ref{fig:binning}.
In addition, we plot the determinant of the parameter covariance matrix as a proxy for the five-dimensional volume.
We observe a quadratic relationship between these metrics and the bin width, with little improvement for bins finer than $\Delta\ell=50$: $\sigma(\Delta\ell=50)/\sigma(\Delta\ell=10) \leq 1.03$ for all parameters.
This is in favour of the choices made by comparable data sets (c.f. \citet{dutcher21, balkenhol23}).

Given that the example likelihood has 25 total parameters (cosmological and nuisance) and deals with covariance matrices with up to $\mathcal{O}(10^6)$ elements, this is no trivial task.
Nevertheless, in the differentiable framework this exercise is easily completed in a few minutes on a laptop, and requires no special structuring of the code to minimise computing costs; this allows for a straightforward translation of ideas into code.
The exercise above can easily be repeated for different covariance matrices based on different survey masks and noise levels, showing how \candl{} can be used to optimise analysis choices.

\subsection{Propagating band power biases to parameters}\label{sec:bdpshift}

Next, we highlight how \candl{} can be used to propagate a bias to band powers $\delta \hat{D}_\ell$ through to parameters $\delta \theta$.
This computation involves the derivative of the model spectra with respect to parameters, $\partial D_\ell / \partial \theta$, and the Fisher matrix, $F$ \citep[see e.g.][]{heavens16}:
\begin{equation}
    \delta \theta = F^{-1} \frac{\partial D_\ell}{\partial \theta} C^{-1} \delta \hat{D}_\ell.
\end{equation}

We use this formalism to reproduce the impact of a $5\%$ rescaling of the \TE{} power spectrum measurement of the ACT DR4 data set.
As shown in \citet{aiola20}, this increase in the \TE{} power leads to a shift in the $\Omega_b h^2 - n_s$ plane.
Using \candl{}, we find a shift down in $\Omega_b h^2$ of $1.6\,\sigma$ and a shift up in $n_s$ of $1.4\,\sigma$.
This appears consistent with what is shown in Figure 14 of \citet{aiola20}.

This result was reproduced within seconds and thanks to automatic differentiation did not require any tuning of step size parameters.
In principle, one can therefore map out the underlying degeneracies in parameter space with ease by rescaling the \TE{} spectrum incrementally.
We note that this approach does not yield an updated parameter covariance, which the re-running of MCMC chains done in \citet{aiola20} produces, though given the comparatively small shift to the best-fit point, we expect that this effect is negligible.\footnote{Though if required, this effect can be approximated by rescaling the parameter covariance matrix obtained from the original MCMC run using the Hessian evaluated at the original and the offset parameter points.}

We note that \citet{mukherjee18} have used this method to obtain parameter constraints from different patches of the \planck{} data.
By propagating the difference between the data band powers and a model spectrum to parameters, one can calculate the best-fit point of the data using an anchor point.
The accuracy of this method depends on the distance between the data best-fit and anchor points.
Applications unlocked by a differentiable pipeline can be used to accomplish a variety of analysis tasks.

\subsection{Parameter-level correlations of correlated band powers}\label{sec:subset_corr}

\begin{figure}
    \centering
    \includegraphics[width=1.0\columnwidth]{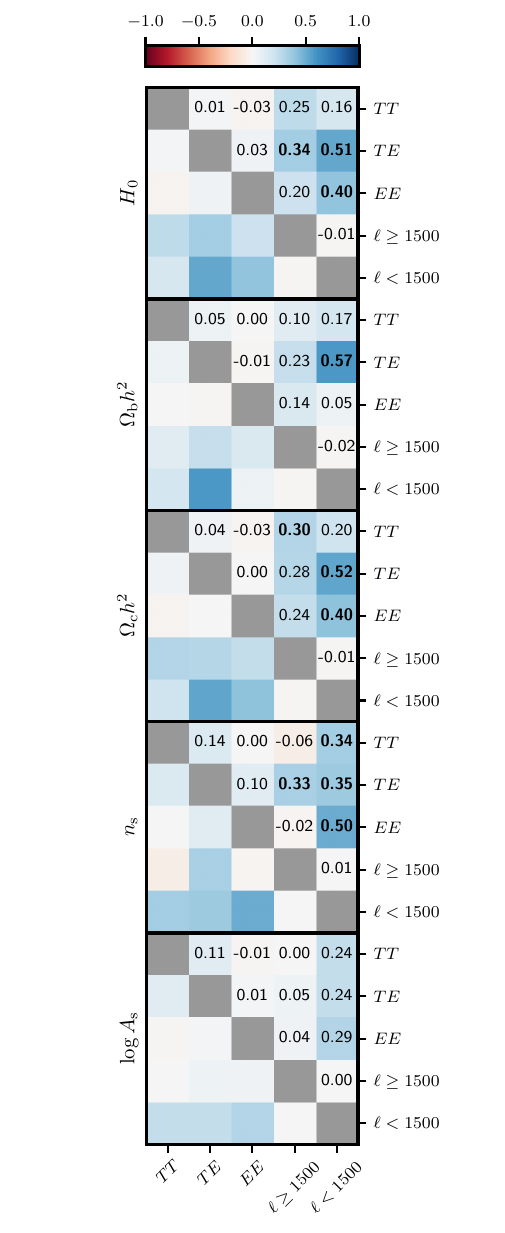}
    \caption{Correlation between best-fit \LCDM{} parameters of the \TT{}, \TE{}, \EE{}, $\ell\geq1500$, and $\ell<1500$ data of SPT-3G 2018 \TTTEEE{} (values $\geq30\%$ in bold).
    We find non-negligible correlation between spectrum types and scale cuts.
    \candl{} makes this calculation simple and robust, enabling a deeper understanding of parameter constraints.}
    \label{fig:par_corr}
\end{figure}

We now use \candl{} and CosmoPower-JAX to calculate the covariance of best-fit parameters of correlated data sets.
We follow the prescription of \citet{kable20}:
\begin{equation}
    \langle ( \bar{\theta}^X - \langle \bar{\theta}^X \rangle ) ( \bar{\theta}^Y - \langle \bar{\theta}^Y \rangle ) \rangle = (M^X)^T C^{XY} M^Y,
\end{equation}
where $\bar{\theta}^{X,Y}$ are the best-fit parameters of data sets $X$ and $Y$, $C^{XY}$ is the relevant block of the band power covariance matrix describing the correlation of $X$ and $Y$ and
\begin{equation}
    M^X = (C^{XX})^{-1} \frac{\partial D_\ell}{\partial \theta} (F^{XX})^{-1},
\end{equation}
where $C^{XX}$ is block the band power covariance matrix for data set $X$ and $F^{XX}$ the Fisher matrix when only considering data set $X$.
The derivatives of the model spectra are typically evaluated at the best-fit point of the full data set \citep{kable20}.

We use the SPT-3G 2018 \TTTEEE{} data set and investigate the following subsets: \TT{}, \TE{}, and \EE{} spectra, and large and small angular scale measurements (split at $\ell = 1500$).\footnote{For this example, we ignore the tSZ-CIB correlation term, as the functional form chosen in \citet{balkenhol23} is not differentiable. Given the tests presented in \S\RNum{4}D therein we expect the impact of this to be negligible.}
On a technical level, the principle task is the calculation of Fisher matrices for all subsets of interest as well as the calculation of the derivatives of the model spectra.

\begin{figure*}
  \centering
  \begin{subfigure}{}
    \includegraphics[width=3.464in]{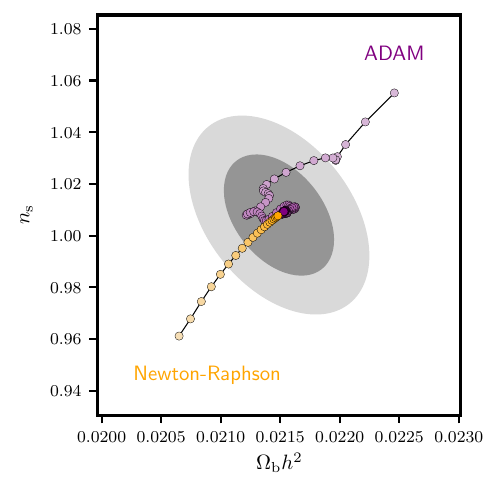}
  \end{subfigure}\hfill
  \begin{subfigure}{}
    \includegraphics[width=3.464in]{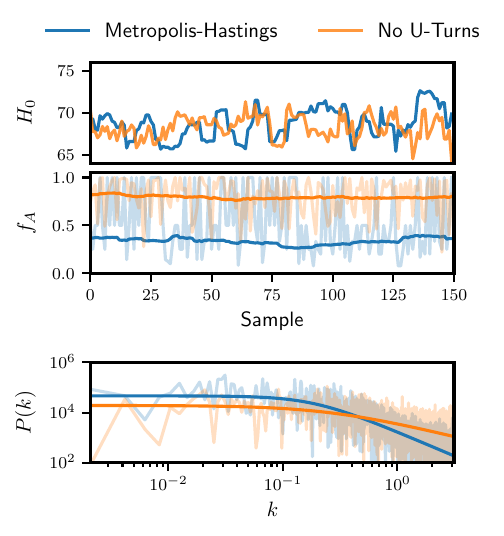}
  \end{subfigure}
  \caption{
Gradient-based explorations of the ACT DR4 likelihood surface in \LCDM{}.\\
\emph{Left:}
Newton-Raphson (orange) and ADAM (purple) minimisers use gradient information to find the best-fit point of the ACT DR4 data set (light to dark points).
We highlight steps in the $n_{\mathrm{s}} - \Omega_{\mathrm{b}} h^2$ plane, though the minimisation is performed over all \LCDM{} parameters and the calibration parameter of the data set.
While the Newton-Raphson algorithm directly descends to the best-fit point, the ADAM minimiser follows a more complicated trajectory with some stochasticity to ensure it does not get trapped in local minima.
The grey contours in the background show the $1$ and $2\,\sigma$ confidence region obtained from the Hessian, which is a by-product of the Newton-Raphson algorithm.\\
\emph{Right:}
Comparison of Metropolis-Hastings (MH) (blue) and No-U-Turns (orange) chains run using the ACT DR4 data set in \LCDM{}.
We focus on $H_0$ samples here, though the MCMC chains explore the full parameter space.
The top and middle panels show the $H_0$ values and acceptance rate $f_A$ (values in faint lines, rolling average over 50 samples in full colour) for an excerpt of $150$ samples, respectively.
The bottom panel shows the power spectrum $P(k)$ of the $H_0$ samples of the full-length chain (measured spectra in faint lines, parametric fit \citep{dunkley05} in full colour).
The NUTS samples are less correlated by eye than the MH samples, which is confirmed by the flatter power spectrum.
Additionally, the acceptance rate of the gradient-based algorithm is substantially higher, with few rejected samples.}
  \label{fig:like_surface}
\end{figure*}

We plot the correlation between the cosmological parameters $H_0$, $\Omega_b h^2$, $\Omega_c h^2$, $n_{\mathrm{s}}$, and $\log{A_{\mathrm{s}}}$ in Figure \ref{fig:par_corr}.
We see that there is little to no correlation between the constraints from different spectra (\TT{}, \TE{}, and \EE{}); the highest correlation we observe is $0.14$ for $n_{\mathrm{s}}$ between \TT{} and \TE{}.
This is testament to the independent information CMB polarisation measurements hold \citep{galli14}.
As expected, constraints from large and small angular scale data are independent of one another; in other words, the off-diagonal elements of each covariance block are small.
While these two results may be expected, there exist non-negligible correlations between these two classes of subsets, which are unknown a priori and only revealed through this calculation.

Access to such correlation matrices facilitates a deeper understanding of the constraints a data set produces in a given model.
A differentiable inference pipeline makes this exercise fast, reliable, and easy.
We note that an alternative approach to calculating the above parameter correlations exists, in which one simulates mock band powers and find their best-fit values; this was done by \citet{kable20} to verify the above procedure.
We note that this approach is also accelerated by \candl{} thanks to access to gradient-based minimisers (see \S\ref{sec:like_exploration}) or using the method of \citet{mukherjee18} described in \S\ref{sec:bdpshift}.

\subsection{Smart exploration of the likelihood surface}\label{sec:like_exploration}

Up until now, we have focussed on applications that are greatly simplified by a differentiable inference pipeline, but in principle with more time, computation power, and caution achievable using finite-difference methods.
We now discuss gradient-based explorations of the likelihood surface, which only become feasible thanks to the easy access to derivatives \candl{} facilitates through JAX's automatic differentiation algorithm.

A common analysis task is determining the minimum of the likelihood, that is the best-fit point of the data.
Gradient-based minimisers allow for a quick descent to the best-fit point and simple algorithms often suffice for Gaussian model spaces when a good starting guess is available.
This is typically the case for primary power spectrum analysis in \LCDM{}, where data is rather constraining and results from previous analyses (e.g. \planck{} \citep{planck18-6}) can be used as a starting point.

As an example, we show the trajectory of two gradient-based minimisers for ACT DR4 \TTTEEE{} in \LCDM{} in the left panel of Figure \ref{fig:like_surface}.
We show a simple Newton-Raphson algorithm \citep[e.g.][]{gill19} alongside the ADAM minimiser \citep{kingma14}.
The former is especially convenient as it requires the computation of the Hessian, which can then be repurposed as the proposal matrix for conventional MCMC chains and short-cut learning this matrix adaptively.
For the ADAM minimiser, we interface the likelihood with 
Optax\footnote{\url{https://github.com/google-deepmind/optax}} \citep{deepmind20}, a JAX-based library of minimisation tools that is widely used in machine-learning.
This algorithm is stochastic and offers protections against getting stuck in local minima.
In this simple example, the Newton-Raphson algorithm and ADAM minimiser perform $20$ and $100$ steps, respectively.
The two algorithms agree to within $\Delta\chi^2 \lesssim 0.2$ and run within seconds.
For reference, the BOBYQA algorithm \citep{powell09, cartis18a, cartis18b} implemented in Cobaya typically requires around $250$ steps for this problem.

In addition to determining the best-fit point of the data, we are often interested in the posterior distribution of parameters, which is typically obtained by building up MCMC samples.
Here, we can use gradient-based MCMC sampling techniques, for example Hamiltonian Monte Carlo \citep{duane87}, Microcanonical Hamiltonian Monte Carlo \citep{robnik22}, or No-U-Turn (NUTS) sampling \citep{hoffman11}.
In these frameworks, the likelihood is typically treated as a potential and the motion of a test mass is modelled.
The result is a higher acceptance rate and a decreased correlation between sample points compared to the Metropolis-Hastings (MH) algorithm; this translates to a better performance in high-dimensional spaces and shorter MCMC chains.

In the right panel of Figure \ref{fig:like_surface}, we show a comparison of the performance of MH and NUTS algorithms for ACT DR4 \TTTEEE{} in \LCDM{}.
The NUTS sampling was done with the aid of BlackJAX\footnote{\url{https://github.com/blackjax-devs/blackjax}} \citep{blackjax20}, a straightforward sampling software written in JAX.
Figure \ref{fig:like_surface} contains an excerpt of $H_0$ samples of MH and NUTS chains; we see by eye that the correlation between neighbouring samples is indeed much lower for the gradient-based algorithm.
We calculate the discrete Fourier transform of the $H_0$ samples for each of the two chains and plot the absolute square of the Fourier coefficients, which corresponds to the power spectrum $P(k)$, in the bottom panel.
We apply the fitting procedure of \citep{dunkley05} to the power spectra.
The fits highlight the suppression of small-scale power for the MH chain due to correlations between close-by point.
In contrast, the NUTS samples resemble more closely the ideal case of uncorrelated white noise.
The power spectrum at $k=0$, $P_0$, of NUTS samples is also lower compared to MH samples ($P^{\mathrm{NUTS}}_0/P^{\mathrm{MH}}_0=0.9$), indicating a higher efficiency and faster convergence \citep{dunkley05, hajian07}.
Lastly, in the example above the proposal matrix of the MH algorithm is set to achieve a close-to optimal acceptance rate of $\sim\!0.25$, meaning that three out of four proposed steps are rejected \citep{roberts97, dunkley05}.
By contrast, the optimal NUTS acceptance rate is $\sim\!0.6$ \citep{hoffman11} and the default target of contemporary implementations is often still higher, for example $0.8$ for BlackJAX and NumPyro \citep{phan19, bingham19, blackjax20}, which is reflected in the example above.

While short chains like in the example above (both chains have around 5000 samples) run in circa one minute on appropriate hardware, for a full MCMC run one must weigh the advantages of different algorithms against their computational costs for the specific problem at hand.
Important aspects to consider are: the size of the data vector,\footnote{We find that the likelihood evaluation time scales quadratically with the length of the data vector.} the number of likelihood evaluations per step, the correlation of samples, the dimensionality of the parameter space, and the available hardware.
One particularly helpful metric in this assessment is the number of effective samples of a chain, which are the number of independent points with the same expression power \citep{whiteley13}.
Running ACT DR4 \LCDM{} chains on CPUs and GPUs, we find that NUTS sampling increases the number of effective samples produced per second by around a factor of four compared to MH sampling.
Put differently, the gradient-based sampler only needs to run for a quarter of the time of the traditional algorithm to achieve the same performance.
We remind the reader that an example notebook covering the three gradient-based algorithms discussed in this section can be found on the GitHub repository.

\subsection{Other applications}\label{sec:other}

Finally, we briefly highlight other applications of \candl{} not discussed so far.
Running example calculations for these suggestions is beyond the scope of this work, though we wish to highlight them here as they stand to benefit from a differentiable likelihood.

A differentiable framework facilitates the construction of compressed likelihoods as was already demonstrated by \citet{campagne23} for the Massively Optimised Parameter Estimation and Data compression \citep{heavens00, zablocki16, heavens17} algorithm.
However, for CMB analysis the more common data compression strategy is to form a set of foreground-marginalised CMB-only band powers from the multifrequency data vector.
This method was put forward by \citet{dunkley13} and has since been applied by the \planck{} and ACT collaborations \citep{planck15-11, louis17, planck18-5, choi20, aiola20}.
Crucially, the framework introduced by \citet{dunkley13} is independent of the cosmological model, such that no theory code (differentiable or otherwise) is required.
One samples directly from the likelihood of the data given a CMB power spectrum and nuisance parameter values.
Typically, this procedure involves alternating Gibbs and MH sampling to build up samples of the CMB band powers and their covariance matrix.

This type of data compression is simplified by \candl{}.
Feeding the likelihood into a Newton-Raphson minimiser appears particularly helpful: the best-fit point yields the correct CMB-only band powers and as a by-product the Hessian provides a good approximation to the covariance matrix.
Following on from this, one can then use a single gradient-based sampling algorithm, such as NUTS, to fully characterise the posterior of CMB-only band powers and obtain their covariance.

Furthermore, easy access to derivatives of the likelihood makes MCMC sampling with the use of the Laplace approximation feasible.
In this approach, one fixes the nuisance parameters to their best-fit values, constructing a so-called profile likelihood.
This reduces the dimensionality of the parameter space to only the parameters of interest, which in turn allows for a faster characterisation of the likelihood surface.
However, one needs to account for the volume reduction as otherwise constraints on cosmological parameters would be artificially tight;
this correction is done by adding a gradient-dependent contribution, the Laplace term, to the obtained $\chi^2$ values.
Thanks to \candl{}, this correction can be computed robustly and sampling accelerated.
We refer the reader to \citet{hadzhiyska23} for a detailed explanation of this method and examples in the cosmology context.

Another application \candl{} seems well suited for is the reconstruction of the primordial power spectrum, $P(k)$, as well as the search for particular features within $P(k)$.
In these types of analyses, one typically adds a large number of additional parameters to the regular cosmological and nuisance parameters, which describe $P(k)$ through one formalism or another \citep{gauthier12, hazra13, handley19, planck13-22}.
For example, a common approach is to model $P(k)$ through $N$ nodes connected by an interpolation scheme and marginalising over the amplitudes and positions of the nodes, hence adding $2N$ parameters.
In this case, traditional MCMC approaches, such as MH sampling, are inefficient, since the required length of chains scales approximately linearly with dimensionality \citep{dunkley05} and the posteriors tend to show complicated features, such as multimodality, which make it difficult to obtain robust constraints \citep{handley19}.
Nested sampling can aid in these scenarios and is helpful for model comparison \citep{skilling06, handley15a, handley15b}, though a fully differentiable pipeline with \candl{} opens the door to further options.
\citet{gauthier12} have already applied a Newton-Raphson minimiser to obtain best-fit solutions in this context;
we can expand on this by using gradient-based sampling methods, which scale better in higher dimensional spaces \citep[e.g.][]{hoffman11}, to fully characterise the posterior.


\section{Getting started}\label{sec:trad}

Lastly, we provide brief instructions for how to get started using \candl{}.
Apart from the final example, the operations below do not require the installation of JAX or a power spectrum emulator;
we refer interested users to the tutorials and in-depth explanations on the GitHub repository.

To initialise a \candl{} likelihood, it suffices to point the code to a data set file or use a short cut for a released data set.
For example:
\begin{lstlisting}[language=Python]
import candl
import candl.data
candl_like = candl.like(candl.data.SPT3G_2018_TTTEEE)
\end{lstlisting}
We can then access all aspects of the data set, such as the band powers, the covariance matrix, the window functions, as well as the data model through attributes of the likelihood instance.

If we have a set of CMB spectra at hand (\lstinline{CMB_specs} below) and want to calculate the corresponding $\chi^2$ value we can proceed with:
\begin{lstlisting}[language=Python]
pars = {'Dl': CMB_specs,
        'nuisance_par': 1.0, ...}
chi2 = candl_like.chi_square(pars)
\end{lstlisting}
where in \lstinline{pars} we pass the CMB spectrum we would like to test, as well as any required nuisance parameter values.

One of the most common use cases for CMB likelihoods is the use with popular sampling software, such as Cobaya.
This is done via the interface module:
\begin{lstlisting}[language=Python]
import candl.interface
from cobaya.run import run
cobaya_info = {'likelihood':
               candl.interface.get_cobaya_info_dict_for_like(candl_like), ...}
updated_cobaya_info = run(cobaya_info)
\end{lstlisting}
The function \lstinline{get_cobaya_info_dict_for_like()} deals with the interface between the two pieces of software.
The user can then select the theory code and sampler they wish and pass these to Cobaya as per usual (see the Cobaya documentation for further details\footnote{\url{https://cobaya.readthedocs.io/en/latest/index.html}}).
A similarly light interface is supplied for MontePython \citep{brinkmann19, audren13}.

Lastly, to calculate Fisher matrices using JAX's automatic differentiation, we assume that we have a differentiable function at hand, \lstinline{pars_to_theory_specs}, that moves from a dictionary of parameter values to CMB spectra.
Helper functions exist in \candl{} to generate such functions based on CosmoPower models, as well as non-differentiable functions based on CAMB, CLASS, Capse.jl, or any \lstinline{Cobaya.Theory} subclass.
Obtaining Fisher matrices is then straightforward with the tools module:
\begin{lstlisting}[language=Python]
import candl.tools
fisher_mat, par_order = candl.tools.get_fisher_matrix(
    pars_to_theory_specs,
    candl_like,
    pars,
    return_par_order = True)
\end{lstlisting}
Here \lstinline{fisher_mat} is the Fisher matrix in the order given by \lstinline{par_order} and we use the same evaluation point (\lstinline{pars}) as before.

\section{Conclusions}\label{sec:conclusions}

In this work, we have presented \candl{}: an automatically differentiable likelihood designed for the analysis of high-precision CMB data.
Through the use of the JAX library, \candl{} offers easy access to derivatives of the likelihood, trivialising in particular the calculation of robust Fisher matrices.
\candl{} is a python-based package that places emphasis on flexibility and efficiency, allowing the user to easily access all aspects of a data set.
The code can easily be interfaced with other established software for instance for MCMC sampling.

We have paired \candl{} with a differentiable theory code, CosmoPower-JAX, and performed a series of example calculations on real and simulated data, showing how the easy access to reliable Fisher matrices opens up new ways of analysing CMB data.
In particular, we have calculated the evolution of parameter errors with respect to the width of band power bins for a mock data set and the correlation of parameter constraints from correlated and overlapping subsets of the SPT-3G 2018 \TTTEEE{} data set.
Moreover, we have showcased gradient-based explorations of the likelihood surface through the use of minimisers and samplers.

The use of differentiable methods in CMB data analysis allows for the construction of a more flexible and reliable pipeline: an increased number of consistency tests and variations on the baseline analysis.
This is crucial for upcoming data sets, which have the potential to surpass \planck{}-precision on cosmological parameters and have the responsibility to deliver a robust analysis.
\candl{} delivers the necessary tools for this task.

We are eager to improve \candl{} in the future.
\candl{} will be used as a part of the upcoming primary anisotropy and lensing analyses of the SPT-3G collaboration;
the associated data sets, as well as those of other CMB experiments, will be made accessible through \candl{} once public.
We are exploring ways of optimising the code to increase its speed.
We are also interested in interfacing the code with probabilistic programming languages, such as NumPyro \citep{phan19, bingham19}.
Lastly, we anticipate the adoption of gradient-based methods by wide-spread cosmology software in the near future and will ensure our likelihood continues to interface with these neatly.

\section*{Acknowledgements}

The authors thank Jo Lauritz for the logo design and are grateful for feedback on \candl{} from \'Etienne Camphuis and the South Pole Telescope Collaboration.
\candl{} uses JAX \citep{jax18} and the scientific python stack \citep{hunter07, jones01, vanDerWalt11}.
This project has received funding from the European Research Council (ERC) under the European Union’s Horizon 2020 research and innovation programme (grant agreement No 101001897).
This work has received funding from the Centre National d’Etudes Spatiales and has made use of the Infinity Cluster hosted by the Institut d’Astrophysique de Paris.
CT was supported by the Center for AstroPhysical Surveys (CAPS) at the National Center for Supercomputing Applications (NCSA), University of Illinois Urbana-Champaign. This work made use of the Illinois Campus Cluster, a computing resource that is operated by the Illinois Campus Cluster Program (ICCP) in conjunction with the National Center for Supercomputing Applications (NCSA) and which is supported by funds from the University of Illinois at Urbana-Champaign.



\bibliographystyle{aa}
\bibliography{candl}


\end{document}